\documentstyle[aps]{revtex}
\begin{document}

\begin{center}
{\bf Deformation of quantum mechanics in fractional-dimensional space}\\

A. Matos-Abiague\footnote{Present address: Max-Planck Institut
f\"{u}r
Mikrostrukturphysik, Weinberg 2, 06120 Halle, Germany \\
Email: amatos@mpi-halle.de \\
Phone: ++49-345-5582-537}\\

Belasitsa 35, Gorna Orjahovitsa 5100, Bulgaria
\end{center}

\vspace{1cm}

\begin{center}
{\bf ABSTRACT}
\end{center}

A new kind of deformed calculus (the D-deformed calculus) that
takes place in fractional-dimensional spaces is presented. The
D-deformed calculus is shown to be an appropriate tool for
treating fractional-dimensional systems in a simple way and quite
analogous to their corresponding one-dimensional partners. Two
simple systems, the free particle and the harmonic oscillator in
fractional-dimensional spaces are reconsidered into the framework
of the D-deformed quantum mechanics. Confined states in a
D-deformed quantum well are studied. D-deformed coherent states
are also found.
\\



\newpage

\section{Introduction}

Fractional-dimensional space approaches have been shown to be
useful in the study of several physical systems. Theoretical
schemes dealing with non-integer space dimensionalities have
frequently been considered in the study of critical phenomena (see
for instance [1], [2]) and of fractal structures [3] or in
modelling semiconductor heterostructure systems [4] - [8].

In the above mentioned schemes, the fractional dimensionality is
not referred to the real space, but to an auxiliary effective
environment used to describe the real system. Nevertheless, the
idea of a real space-time having a dimension slightly different
from four has also been considered by several authors [9] - [12].
Actually, the deviations of the space-time dimension from four
have been found to be very small [9] - [12]. However, the question
of whether the dimension of the space-time is an integer or a
fractional number constitutes a basic problem not only for its
conceptual significance but also because the possibility that the
space-time dimension is different from four may lead to
interesting consequences (e.g. it is well known that a deviation
of the space-time dimension from the value four eliminates the
logarithmic divergences of quantum electrodynamics, independently
of how small the deviation from four may be [13]).

Recently, it has been shown the existence of some similarities
between the so-called N-body Calogero models and the problem
corresponding to a fractional-dimensional harmonic oscillator of a
single degree of freedom [14]. This result together with the
remarkable fact that fractional-dimensional bosons can be
considered as generalized parabosons [14] suggests new potential
applications of the non-integer-dimensional space approaches.

It was shown in [14] that the fractional-dimensional Bose
operators together with the reflection operator form an R-deformed
Heisenberg algebra with a deformation parameter depending on the
dimension of the space. Deformations of the Heisenberg algebra
leading to the so-called q-deformed quantum mechanics have been
extensively investigated (see for instance [15] - [18]). Taking
into account the results obtained in [14], we develop in the
present paper a new deformed calculus (the D-deformed calculus) in
analogy to the q-deformed calculus commonly treated in the
literature [18] - [21]. The new calculus allow us to express and
to solve problems concerning fractional-dimensional systems in a
simple way and quite analogous to the corresponding undeformed
($D=1$) problems. The paper is organized as follows. In section 2
we introduce the D-deformed calculus. The problems corresponding
to the free particle and to the harmonic oscillator in
fractional-dimensional space were studied in [14]. We reconsider
these problems in sections 3 and 4 respectively, but now from the
point of view of the D-deformed quantum mechanics. Of course, the
final results in these sections coincide with the results obtained
in [14]. However, in terms of the new D-deformed calculus, the
above mentioned problems can be solved immediately and in an
elegant way. In section 5 the single particle confined states in a
fractional-dimensional quantum well are studied. The dimensional
dependence of the eigenenergies corresponding to the ground and to
the first excited states is shown. In both cases an increase of
the energies as the dimension increases is observed. The
probability density function describing the motion of the particle
confined in the D-deformed quantum well is also studied for
varying the dimensionality. The D-deformed coherent states are
found in section 6 and conclusions are summarized in section 7.

\section{D-deformed calculus}

It is well known that the one-dimensional momentum operator is given by

\begin{equation}
P=\frac 1i\frac d{d\xi }\quad ,  \label{e2.1}
\end{equation}

where we have taken $\hbar =1$. However, in a
fractional-dimensional space, because of the inclusion of the
integration weight [22]

\begin{equation}
\frac{\sigma (D)}2\left| \xi \right| ^{D-1},\qquad \sigma (D)=\frac{2\pi
^{D/2}}{\Gamma (D/2)}\quad ,  \label{e2.2}
\end{equation}

this operator is no longer Hermitian. Therefore a more general
momentum operator has to be defined for systems in
fractional-dimensional spaces. Starting with the Wigner
commutation relations for the canonical variables of a Bose-like
oscillator of a single degree of freedom, the
fractional-dimensional momentum operator has been found to be
[14],

\begin{equation}
P=\frac 1i\frac d{d\xi }+i\frac{(D-1)}{2\xi }R-i\frac{(D-1)}{2\xi }\quad ,
\label{e2.3}
\end{equation}

where $R$ is the reflection operator.

The momentum operator (equation (\ref{e2.3})) suggests a
deformation of quantum mechanics in fractional-dimensional spaces.
Indeed, we can introduce a new D-deformed derivative operator

\begin{equation}
\frac{d_D}{d_D\xi }=\frac d{d\xi }+\frac{(D-1)}{2\xi }(1-R)\quad
\label{e2.4}
\end{equation}

and then the fractional-dimensional momentum operator (equation
(\ref{e2.3})) can be rewritten in the standard form

\begin{equation}
P=\frac 1i\frac{d_D}{d_D\xi }\quad .  \label{e2.5}
\end{equation}

Thus the D-deformed annihilation and creation operators can be defined in
the following way

\begin{equation}
a_D=\frac 1{\sqrt{2}}\left( \xi +\frac{d_D}{d_D\xi }\right) \quad
;\quad a_D^{\dagger }=\frac 1{\sqrt{2}}\left( \xi -
\frac{d_D}{d_D\xi }\right) \quad .  \label{e2.6}
\end{equation}

The action of these operators is given as follows [14]

\begin{equation}
a_D|0\rangle =0\quad ,  \label{e2.7}
\end{equation}

\begin{equation}
a_D|2n\rangle =\sqrt{2n}|2n-1\rangle \quad ;\quad a_D|2n+1\rangle =\sqrt{2n+D%
}|2n\rangle \quad ,  \label{e2.8}
\end{equation}

\begin{equation}
a_D^{\dagger }|2n\rangle =\sqrt{2n+D}|2n+1\rangle \quad ;\quad a_D^{\dagger
}|2n+1\rangle =\sqrt{2n+2}|2n+2\rangle \quad ,  \label{e2.9}
\end{equation}

where $n=0,1,2,3,...$

By now introducing the corresponding D-factor (analogue to the q-factor) as

\begin{equation}
\lbrack n]_D=n+\frac{(D-1)}2(1-(-1)^n)\quad ,  \label{e2.10}
\end{equation}

the equations (\ref{e2.8}) and (\ref{e2.9}) can be rewritten in
the usual form

\begin{equation}
a_D|n\rangle =\sqrt{[n]_D}|n-1\rangle \quad ,  \label{e2.11}
\end{equation}

and

\begin{equation}
a_D^{\dagger }|n\rangle =\sqrt{[n+1]_D}|n+1\rangle \quad ,  \label{e2.12}
\end{equation}

respectively.

Taking into account equation (\ref{e2.10}) and in analogy to the
q-deformed standard procedures we can define a D-deformed
factorial function as follows

\begin{equation}
\lbrack n]_D!=[n]_D[n-1]_D...[1]_D[0]_D!=\left\{
\begin{array}{c}
\ \ \frac{2^n\left( \frac n2\right) !\Gamma (\frac{n+D}2)}{\Gamma (D/2)}%
\qquad \qquad \ \text{for n even \ } \\
\frac{2^n\left( \frac{n-1}2\right) !\Gamma (\frac{n+D+1}2)}{\Gamma (D/2)}%
\qquad \ \ \ \ \text{for n odd }
\end{array}
\right. \quad .  \label{e2.13}
\end{equation}

This D-deformed factorial function is a particular case of the
generalized factorial function [23].

The eigenstates $|n\rangle $ of the operator

\begin{equation}
N_D|n\rangle =n|n\rangle \quad ;\quad N_D=\frac 12\{a_D^{\dagger },a_D\}-D/2
\label{e2.14}
\end{equation}

may be obtained by repeated applications of $a_D^{\dagger }$ on the vacuum
state $|0\rangle $

\begin{equation}
|n\rangle =\frac{\left( a_D^{\dagger }\right) ^n}{\sqrt{[n]_D!}}|0\rangle
\quad .  \label{e2.15}
\end{equation}

It is easy to prove that in this Fock space, the relations

\begin{equation}
a_D^{\dagger }a_D=[n]_D\quad ;\quad a_Da_D^{\dagger }=[n+1]_D  \label{e2.16}
\end{equation}

take place.

From the definition of the D-deformed derivative (equation
(\ref{e2.4})) we can introduce a D-deformed integration, so that
if

\begin{equation}
\frac{d_Df(\xi )}{d_D\xi }=F(\xi )\quad ,  \label{e2.17}
\end{equation}

then

\begin{equation}
f(\xi )=\int F(\xi )d_D\xi +const.  \label{e2.18}
\end{equation}

With the aim to find the appropriate expression for the D-deformed
integration, we observe that

\begin{equation}
\frac{d_Df(\xi )}{d_D\xi }=\left[ 1+\frac{(D-1)}{2\xi }(1-R)\int d\xi
\right] \frac{df(\xi )}{d\xi }=F(\xi )\quad ,  \label{e2.19}
\end{equation}

and hence

\begin{equation}
\frac{df(\xi )}{d\xi }=\left[ 1+\frac{(D-1)}{2\xi }(1-R)\int d\xi \right]
^{-1}F(\xi )\quad .  \label{e2.20}
\end{equation}

From the equation above it follows

\begin{equation}
f(\xi )=\int F(\xi )d_D\xi =\sum\limits_{n=0}^\infty \left[ -\int d\xi \frac{%
(D-1)}{2\xi }(1-R)\right] ^n\int d\xi F(\xi )\quad .  \label{e2.21}
\end{equation}

This expression may be rewritten as

\begin{equation}
\int F(\xi )d_D\xi =\sum\limits_{n=0}^\infty (-1)^nI_n\quad ,  \label{e2.22}
\end{equation}

where the terms $I_n$ satisfy the following recurrence formula

\begin{equation}
I_{n+1}=\int \frac{(D-1)}{2\xi }(1-R)I_nd\xi \quad ;\quad %
I_0=\int F(\xi )d\xi \quad .  \label{e2.23}
\end{equation}

With respect to the D-deformed calculus induced by the
fractional-dimensional integration weight (equation (\ref{e2.2})),
the following identities can be easily demonstrated

\begin{equation}
\frac{d_D[f(\xi )g(\xi )]}{d_D\xi }=g(\xi )\frac{d_Df(\xi )}{d_D\xi }+\frac{%
d_Dg(\xi )}{d_D\xi }Rf(\xi )+\frac{dg(\xi )}{d\xi }(1-R)f(\xi )
\label{e2.24}
\end{equation}

and after integrating the equation above

\begin{equation}
\int g(\xi )\frac{d_Df(\xi )}{d_D\xi }d_D\xi =f(\xi )g(\xi )-\int \frac{%
d_Dg(\xi )}{d_D\xi }Rf(\xi )d_D\xi -\int \frac{dg(\xi )}{d\xi }(1-R)f(\xi
)d_D\xi \quad .  \label{e2.25}
\end{equation}

One should notice that if either $f(\xi )$ or $g(\xi )$ is an even
function of $\xi $, equations (\ref{e2.24}) and (\ref{e2.25})
reduce to a D-deformed Leibnitz rule and to a D-deformed formula
of integration by parts, respectively. This is a consequence of
the fact that the D-deformed derivative acts on even functions as
the ordinary derivative.

\section{D-deformed free particle}

The eigenstates of the momentum operator corresponding to a free particle of
a single degree of freedom in a fractional-dimensional space can be found
now in terms of the D-deformed calculus introduced in the previous section.
Thus, the eigenstates of the fractional-dimensional momentum operator are
determined by the following equation

\begin{equation}
P\Psi _p=-i\frac{d_D\Psi _p}{d_D\xi }=p\Psi _p\quad .
\label{e3.1}
\end{equation}

The corresponding eigenfunctions are immediately found to be

\begin{equation}
\Psi _p=A_pE_D(ip\xi )\quad ,  \label{e3.2}
\end{equation}

where $E_D(x)$ represents the D-deformed exponential function (see Appendix
A). The normalization factor $A_p$ can be found from the orthonormalization
condition

\begin{equation}
\langle \Psi _p|\Psi _{p^{\prime }}\rangle =\frac{\sigma (D)}2%
\lim\limits_{\gamma \rightarrow 0}\int_{-\infty }^\infty
e^{-\gamma \xi ^2}\Psi _p^{*}(\xi )\Psi _{p^{\prime }}(\xi )|\xi
|^{D-1}d\xi =\delta (p-p^{\prime })\qquad (\gamma >0),
\label{e3.3}
\end{equation}

in a similar way as in [14]. After the corresponding calculations
we arrive to the following expression

\begin{equation}
A_p=\frac 1{2^{D/2-1}\Gamma (D/2)}\sqrt{\frac{p^{D-1}}{2\sigma (D)}}\quad .
\label{e3.4}
\end{equation}

One should notice that the eigenfunctions $\Psi _p$ describing the motion of
a free particle of a single degree of freedom in a fractional-dimensional
space can be considered as D-deformed plane waves and they reduce to the
ordinary plane de Broglie waves when $D=1$.

\section{D-deformed harmonic oscillator}

The eigenfunctions in coordinate representation corresponding to
the D-deformed harmonic oscillator can be derived from equation
(\ref{e2.15}) without much difficulty. First we consider the
vacuum state $|0\rangle $ which satisfies equation (\ref{e2.7}).
Then, using the expression of $a_D$ in coordinate representation
(equation (\ref{e2.6})) we have the following D-deformed
differential equation

\begin{equation}
\left( \frac{d_D}{d_D\xi }+\xi \right) \chi _0=0\quad ,  \label{e4.1}
\end{equation}

where $\chi _0=\langle \xi |0\rangle $ represents the
eigenfunction of the ground state of the D-deformed harmonic
oscillator. By now solving equation (\ref {e4.1}) we found that

\begin{equation}
\chi _0=C_0\exp [-\xi ^2/2]\quad ,  \label{e4.2}
\end{equation}

where $C_0$ is a normalization factor. From the normalization condition

\begin{equation}
\frac{\sigma (D)}2\int\limits_{-\infty }^\infty |\chi _0|^2|\xi
|^{D-1}d\xi =1\quad ,  \label{e4.3}
\end{equation}

the normalization constant is found to be

\begin{equation}
C_0=\frac 1{\pi ^{D/4}}\quad .  \label{e4.4}
\end{equation}

Once the ground state has been found, the excited states may be
calculated from equation (\ref{e2.15}). Thus the excited states
are determined by

\begin{equation}
\chi _n=\langle \xi |n\rangle =\frac{C_0}{\sqrt{[n]_D!}}\left[ \frac 1{\sqrt{%
2}}\left( \xi -\frac{d_D}{d_D\xi }\right) \right] ^n\exp [-\xi ^2/2]\quad .
\label{e4.5}
\end{equation}

If we now take into account that

\begin{equation}
(-1)^n\exp [\xi ^2/2]\left( \frac{d_D}{d_D\xi }\right) ^n\exp [-\xi
^2]=\left( \xi -\frac{d_D}{d_D\xi }\right) ^n\exp [-\xi ^2/2]\quad ,
\label{e4.6}
\end{equation}

a relation that can be demonstrated by induction, the excited
states in coordinate representation can be written as

\begin{equation}
\chi _n=\langle \xi |n\rangle =\frac{[n]_D!\exp [-\xi ^2/2]}{n!\sqrt{\pi
^{D/2}2^n[n]_D!}}H_n^D(\xi )\quad ,  \label{e4.7}
\end{equation}

where

\begin{equation}
H_n^D(\xi )=\frac{n!}{[n]_D!}(-1)^n\exp [\xi ^2]\left( \frac{d_D}{d_D\xi }%
\right) ^n\exp [-\xi ^2]\quad ,  \label{e4.8}
\end{equation}

can be understood as D-deformed Hermite polynomials. In fact the
polynomials $H_n^D(\xi )$ above defined are a particular case of
the generalized Hermite polynomials studied in [23].

It is worth remarking that if $D=1$ the results obtained in the present
section reduce to the well known results corresponding to the {\it undeformed%
} one-dimensional case.

\section{Confined states in a D-deformed quantum well}

In the present section we will study the motion of a particle
confined in a fractional-dimensional quantum well defined by the
following potential

\begin{equation}
V(\xi )=\left\{
\begin{array}{c}
0 \quad \text{if }|\xi|< 1/2 \\
\infty \quad \text{otherwise }
\end{array}
\right. ,  \label{e5.1}
\end{equation}

where we have taken an unitary well width ($L=1$). The
corresponding Schr\"{o}dinger equation may be written as

\begin{equation}
\left[-\frac{1}{2}\frac{d_D^2}{d_D\xi^{2}}+V(\xi)\right]\Psi_n(\xi)=E_n
\Psi_n(\xi)\quad . \label{e5.2}
\end{equation}

In terms of the introduced D-deformed calculus, the equation above
can be immediately solved. The wavefunctions are given by

\begin{equation}
\Psi_n(\xi )=\left\{
\begin{array}{c}
A_n^{even}COS_{D} k_n\xi \quad \text{for n even } \\
A_n^{odd}SIN_{D} k_n\xi \quad \text{for n odd }
\end{array}
\right. ,  \label{e5.3}
\end{equation}

where

\begin{equation}
k_n=\sqrt{2E_n} \quad , \label{e5.4}
\end{equation}

and $COS_{D}X$, $SIN_{D}X$ represent the D-deformed
\textit{cosine} and \textit{sine} functions respectively (see
Appendix). The constants $A_n^{even}$, $A_n^{odd}$ are
normalization factors corresponding to even and odd states
respectively.

The eigenenergies can be now easily computed from the boundary
conditions

\begin{equation}
COS_D\frac{k_n}{2}=0 \quad , \label{e5.5}
\end{equation}

for even states and

\begin{equation}
SIN_D\frac{k_n}{2}=0 \quad , \label{e5.6}
\end{equation}

for odd states.

One should notice that the D-deformed calculus allow us to express
the fractional-dimensional problems in a very simple way and quite
analogous to the corresponding undeformed ($D=1$) problems.

The D-dependence of the energies corresponding to the ground state
($n=0$) and to the first excited state ($n=1$) is shown in figure
1. The eigenenergies increase as the dimension increases. This
behaviour has also been observed in other fractional-dimensional
systems\footnote{The energies of the fractional-dimensional
harmonic oscillator and the hydrogenic atom are given by
$E_{n}=n+D/2$ and $E_{n}=\frac{-4}{(2n+D-3)^2}$, respectively. In
both cases the energy increases when the dimension increases.}
(see for instance [4], [14]).

In figure 2 we present the probability density

\begin{equation}
\rho_{n}(\xi)=\frac{\sigma(D)}{2}|\xi|^{D-1}|\Psi_{n}|^2 \quad ,
\label{e5.7}
\end{equation}

corresponding to $n=0$ (a) and to $n=1$ (b) as a function of the
pseudocoordinate $\xi$ for different values of the dimensionality.
In both cases one can appreciate the existence of compression
(spreading) of the probability density when $D<1$ ($D>1$). We
remark that this behaviour is not a consequence of the form of the
wavefunctions but of the presence of the integration weight in the
probability density. Indeed, the integration weight acts as an
attractive (repulsive) barrier when $D<1$ ($D>1$). It diverges at
the origin of pseudocoordinates when $D<1$ (i. e. when $D<1$ all
the volume of the space is almost concentrated around $\xi=0$ )
and causes a strong localization of the probability density (see
figure 2(a)). In the case of the first excited state, however,
because of the odd parity the probability density becomes zero at
the origin of pseudocoordinate and there is no longer localization
in the central region.

It is worth noting that in the present case when $D>1$, the
maximum of the probability density increases as the dimension
increases (see figure 2), contrary to the behaviour observed in
the fractional-dimensional harmonic oscillator (see figures 2(b)
and 3 of [14]). This is because as the dimensionality increases,
the integration weight becomes more and more repulsive favouring
the tunnelling through the harmonic barriers. Consequently, the
particle becomes more delocalized and the maximum of the
probability density decreases. In the present case, however, the
quantum well is considered infinitely deep and the tunnelling is
suppressed. The particle is then \textit{compressed} between the
well barriers and the repulsive integration weight leading to an
increase in the maximum of the probability density.

\section{D-deformed coherent states}

We now observe the spectrum problem corresponding to a
fractional-dimensional annihilation operator $a_D$ by using the
rules of the D-deformed calculus. The eigenstates of $a_D$:

\begin{equation}
a_D|\alpha \rangle =\alpha |\alpha \rangle  \label{e6.1}
\end{equation}

are a D-deformation of the usual coherent states. The solution of equation (\ref{e6.1}%
) is given by

\begin{equation}
|\alpha \rangle =A_\alpha \sum\limits_{n=0}^\infty \frac{\alpha ^n}{\sqrt{%
[n]_D!}}|n\rangle \quad .  \label{e6.2}
\end{equation}

From the normalization condition $\langle \alpha |\alpha \rangle=1
$ the normalization constant is found to be

\begin{equation}
A_\alpha =\frac 1{\sqrt{E_D(|\alpha |^2)}} \quad .  \label{e6.3}
\end{equation}

As usually, the equation (\ref{e6.2}) can be written in term of
the vacuum state as follows

\begin{equation}
|\alpha \rangle =\frac 1{\sqrt{E_D(|\alpha |^2)}}E_D(\alpha
a_D^{\dagger })|0\rangle \quad .  \label{e6.4}
\end{equation}

Once we have found the expression of the D-deformed coherent states in the
Fock representation, we can easily obtain its expression in coordinate
representation by using the relation

\begin{equation}
\Phi _\alpha (\xi )=\langle \xi |\alpha \rangle
=\sum\limits_{n=0}^\infty \langle \xi |n\rangle \langle n|\alpha
\rangle \quad .  \label{e6.5}
\end{equation}

By now considering equations (\ref{e4.7}) and (\ref{e6.2}) we
arrive to the following result

\begin{equation}
\Phi _\alpha (\xi )=\frac{\exp [-\xi ^2/2]}{\sqrt{E_D(|\alpha |^2)}}\frac 1{%
\pi ^{D/4}}\sum\limits_{n=0}^\infty \frac{\alpha
^n}{\sqrt{2^n}n!}H_n^D(\xi )\quad .  \label{e6.6}
\end{equation}

From equation (\ref{e6.2}), the probability distribution of a
D-deformed coherent state in Fock representation is found to be

\begin{equation}
|\langle n|\alpha \rangle |^2=\frac 1{E_D(|\alpha |^2)}\frac{(|\alpha |^2)^n%
}{[n]_D!}  \label{e6.7}
\end{equation}

i. e. a D-deformation of the Poisson distribution.

\section{Conclusions}

Summing up, taking into account recent developments in the
mathematical physics of the fractional-dimensional space and in
analogy to the q-deformed calculus we have developed a new
deformed calculus that we have called D-deformed calculus. Some
simple fractional-dimensional systems, the free particle, the
harmonic oscillator, and the particle confined in a quantum well
have been studied into the framework of the D-deformed quantum
mechanics. Finally, the D-deformed coherent states are found.

\appendix

\section*{}

Here we will study the properties of some D-deformed functions.
From the definition of the D-deformed derivative (equation
(\ref{e2.4})) it is easy to find that

\begin{equation}
\frac{d_D\xi ^n}{d_D\xi }=[n]_D\xi ^{n-1}\quad .  \label{A1}
\end{equation}

On the other hand, from the definition of the D-deformed integration (equation (%
\ref{e2.22})), we also found

\begin{equation}
\int \xi ^nd_D\xi =\frac{\xi ^{n+1}}{[n+1]_D}+const.  \label{A2}
\end{equation}

In this way, one can introduce the D-deformed exponential function as follows

\begin{equation}
E_D(\xi )=\sum\limits_{n=0}^\infty \frac{\xi ^n}{[n]_D!} \quad .
\label{A3}
\end{equation}

From equation (\ref{A3}) and making use of the equations
(\ref{A1}) and (\ref{A2}) one can straightforwardly demonstrate
that

\begin{equation}
\frac{d_DE_D(\lambda \xi )}{d_D\xi }=\lambda E_D(\lambda \xi )\quad ;\quad
\lambda =const.  \label{A4}
\end{equation}

and consequently

\begin{equation}
\int E_D(\lambda \xi )d_D\xi =\frac{E_D(\lambda \xi )}\lambda +const.
\label{A5}
\end{equation}

Actually, the D-deformed exponential function is a particular case
of the generalized exponential function defined in [23] and can be
represented as follows

\begin{equation}
E_D(\xi )=\exp [\xi ]\Phi (\frac{D-1}2,D,-2\xi )\quad ,  \label{A6}
\end{equation}

or

\begin{equation}
E_D(\xi )=\Gamma (D/2)\left( \frac \xi 2\right) ^{1-D/2}\left[ I_{D/2-1}(\xi
)+I_{D/2}(\xi )\right] \quad ,  \label{A7}
\end{equation}

where $\Phi(a,b,x) $ and $I_\nu (x) $ are the confluent
hypergeometric function and the modified Bessel function
respectively.

We can also introduce D-deformed {\it cosine} and {\it sine} functions
through the following definitions

\begin{equation}
COS_D\xi =\sum\limits_{n=0}^\infty \frac{(-1)^n\xi ^{2n}}{[2n]_D!}=\Gamma
(D/2)\left( \frac \xi 2\right) ^{1-D/2}J_{D/2-1}(\xi )\quad ,  \label{A8}
\end{equation}

and

\begin{equation}
SIN_D\xi =\sum\limits_{n=1}^\infty \frac{(-1)^{n-1}\xi ^{2n-1}}{[2n-1]_D!}%
=\Gamma (D/2)\left( \frac \xi 2\right) ^{1-D/2}J_{D/2}(\xi )\quad ,
\label{A9}
\end{equation}

where $J_\nu (x)$ represents the Bessel function. Thus, the
following identities can be easily verified

\begin{equation}
E_D(\pm i\xi )=COS_D\xi \pm i\,SIN_D\xi \quad ,  \label{A10}
\end{equation}

and

\begin{equation}
\frac{d_DCOS_D\xi }{d_D\xi }=-SIN_D\xi \quad ;\quad \frac{d_DSIN_D\xi }{%
d_D\xi }=COS_D\xi  \label{A11}
\end{equation}

It is straightforwardly to check that all the definitions and
equations given in this appendix recover the corresponding
undeformed expressions when $D=1$, as it must.

\begin{center}
{\bf References }
\end{center}

\begin{enumerate}
\item  Ma S 1973 \textit{Rev. Mod. Phys.} {\bf 45} 589

\item  Fisher M E 1974 \textit{Rev. Mod. Phys.} {\bf 46} 597

\item  Mandelbroot B B 1989 {\it The Fractal Geometry of Nature} (San
Francisco: Freeman)

\item  He X F 1991 \textit{Phys. Rev.} B {\bf 43} 2063

\item  Mathieu H, Lefebvre P and Christol P 1992 \textit{Phys. Rev.} B {\bf
46} 4092

\item  Christol P, Lefebvre P and Mathieu H 1993 \textit{J. Appl. Phys}. {\bf 74}
5626

\item  Matos-Abiague A, Oliveira L E and de Dios-Leyva M 1998 \textit{Phys.
Rev.} B {\bf 58} 4072

\item  Reyes-G\'omez E, Matos-Abiague A, Perdomo-Leiva C A, de
Dios-Leyva M and Oliveira L E 2000 \textit{Phys. Rev.} B {\bf 61}
13104

\item  Zeilinger A and Svozil K 1985 \textit{Phys. Rev. Lett.} {\bf 54} 2553

\item  Jarlskog C and Yndur\'ain F J 1986 E\textit{urophys. Lett.} {\bf 1} 51

\item  Sch\"afer A and M\"uller B 1986 \textit{J. Phys. A: Math. Gen.} {\bf 19}
3981

\item  Torres J L and Ferreira Herrej\'on P 1989, \textit{Rev. Mex. F\'\i s.} {\bf 35}
97

\item  Weisskopf V F 1939 \textit{Phys. Rev.} {\bf 56} 72

\item  Matos-Abiague A 2001 \textit{J. Phys. A: Math. Gen.} {\bf 34} 3125

\item  Macfarlane A J 1989 \textit{J. Phys. A: Math. Gen.} {\bf 22} 4581

\item  Sun C P and Fu H C 1989 \textit{J. Phys. A: Math. Gen.} {\bf 22} L983

\item  Chang Z, Chen W, Guo H Y and Yan H 1990 \textit{J. Phys.} A {\bf 23}
5371

\item  Biedenharn L C 1989 \textit{J. Phys. A: Math. Gen.} {\bf 22} L873

\item  Li Y and Sheng Z 1992 \textit{J. Phys. A: Math. Gen.} {\bf 25} 6779

\item  Chung W, Chung K, Nam S and Um C 1993 \textit{Phys. Lett.} {\bf 183A}
363

\item  Chung K and Chung W 1994 \textit{J. Phys. A: Math. Gen.} {\bf 27} 5037

\item  Stillinger F H 1977 \textit{J. Mat. Phys.} {\bf 18} 1224

\item  Rosenblum M 1994 {\it Operator Theory: Advances and Applications}
vol. 73 (Basel: Birkh\"auser) pp 369-396

\end{enumerate}

\newpage

{\bf Figure captions}

Figure 1. The eigenenergies corresponding to the ground ($n=1$)
and to the first exited ($n=2$) states of a particle confined in a
D-deformed quantum well as a function of the dimensionality.

Figure 2. Position dependence of the probability density
$\rho_{n}$ corresponding to a particle confined in a D-deformed
quantum well and for different values of the dimensional
parameter. (a) For the ground state ($n=1$) and (b) for the first
exited state ($n=2$).

\end{document}